\documentclass[a4paper,justified,notitlepage,nobib]{tufte-handout}

\usepackage{amsmath,amssymb,amsthm}
\usepackage{tikz}
\usetikzlibrary{arrows.meta,positioning,shapes.geometric,
                fit,backgrounds,decorations.pathreplacing,
                matrix,calc}
\usepackage{nicefrac}
\usepackage{booktabs}
\usepackage{microtype}
\usepackage{xcolor}
\usepackage{algorithm}
\usepackage{algpseudocode}
\usepackage{pgfplots}
\pgfplotsset{compat=1.18}
\usepgfplotslibrary{groupplots}
\usepackage[numbers,sort&compress]{natbib}  % numeric [1,2] citations
\usepackage{listings}
\usepackage[most]{tcolorbox}

\definecolor{pyblue}{RGB}{31,119,180}       % Python — blue
\definecolor{pybg}{RGB}{235,244,255}
\definecolor{julgreen}{RGB}{39,139,58}      % Julia  — green
\definecolor{julbg}{RGB}{235,250,238}
\definecolor{cppred}{RGB}{180,40,40}        % C++    — red
\definecolor{cppbg}{RGB}{255,238,238}
\definecolor{outgray}{RGB}{60,60,60}        % output — dark gray
\definecolor{outbg}{RGB}{245,245,245}
\definecolor{codegray}{RGB}{100,100,100}

\tcbset{
  codebox/.style={
    fonttitle=\bfseries\small,
    colbacktitle=#1!80!black,
    colback=#1!10!white,
    coltitle=white,
    colframe=#1!60!black,
    boxrule=0.6pt,
    arc=3pt,
    left=4pt, right=4pt, top=2pt, bottom=2pt,
    listing only,
    listing options={basicstyle=\ttfamily\scriptsize, breaklines=true,
                     columns=flexible, keepspaces=true},
  },
}

\usepackage[skip=8pt, font=small]{caption}

\definecolor{tuftered}{RGB}{179,27,27}
\definecolor{tufteblue}{RGB}{31,73,125}
\definecolor{tuftegreen}{RGB}{0,110,60}

\newtheoremstyle{tuftethm}{}{}{\itshape}{}{\bfseries}{.}{.5em}{}
\theoremstyle{tuftethm}
\newtheorem{definition}{Definition}

\newcommand{\D}{\mathsf{D}}
\newcommand{\N}{\mathsf{N}}
\newcommand{\Qmat}{\mathbf{Q}}
\newcommand{\Emat}{\mathbf{E}}
\newcommand{\Rmat}{\mathbf{R}}
\newcommand{\bx}{\mathbf{x}}
\newcommand{\oct}[1]{\ensuremath{#1_8}}

\title{Distance Spectrum of IEEE 802.11 Binary Convolutional Codes}
\author{Rethna Pulikkoonattu\\\small\nolinkurl{rethna@broadcom.com}}
\date{}

\begin{document}
\maketitle

\begin{abstract}
\noindent
Binary convolutional coding (BCC) has been a cornerstone of the IEEE~802.11
wireless LAN standard since its inception, and it remains relevant today across
the full generational arc from the legacy \mbox{802.11a/g} through Wi-Fi~6
(802.11ax) and into the forthcoming Wi-Fi~8 (802.11bn)~\cite{ieee80211,ieee80211bn}.
Although low-density parity-check (LDPC) codes now dominate high-throughput
applications, BCC is mandatory for backward compatibility and continues to serve
as the \emph{default} forward-error-correction scheme in bandwidth-constrained
and cost-sensitive deployments: 20\,MHz-only devices, Internet-of-Things nodes,
and other implementations where LDPC's decoder complexity is prohibitive.
Critically, BCC at rate~\nicefrac{1}{2} is the coding scheme used throughout
the packet preamble in every IEEE~802.11-compliant frame, making it indispensable
regardless of which data-field code is selected.  Furthermore, the new
\emph{Enhanced Long Range} (ELR) packet format introduced in the
802.11bn/UHR amendment mandates rate-\nicefrac{1}{2} BCC for the data
portion of the frame, reinforcing the continued importance of this code in
next-generation deployments~\cite{ieee80211bn}.

The performance of BCC under Viterbi decoding is governed by the
\emph{distance spectrum} $\{(\alpha_d,\beta_d)\}_{d\geq d_\mathrm{free}}$
of the convolutional code~\cite{lin04,viterbi67}.
This note explains how to compute that spectrum
exactly for the IEEE~802.11 mother code (rate~\nicefrac{1}{2}, $K=7$,
generators $\oct{133}$/$\oct{171}$) and its three standard punctured derivatives
(rates \nicefrac{2}{3}, \nicefrac{3}{4}, \nicefrac{5}{6}) obtained via
rate-compatible puncturing~\cite{cain79,hagenauer88}.
Union-bound BEP and FER curves are derived for AWGN with BPSK/QPSK and
Gray-coded $M$-QAM modulation and validated against Monte Carlo simulation.
Python, Julia, and C++ implementations are openly available at~\cite{bcc_github}.
\end{abstract}

%% =======================================================
%% =======================================================
\section{Introduction}
%% =======================================================

Despite the ubiquity of BCC in deployed IEEE~802.11 systems, a recurring
obstacle when analysing or benchmarking its performance is the absence of
a self-contained, publicly accessible resource that derives the distance
spectrum from first principles and connects it cleanly to BER and FER
bounds.  Existing textbook treatments~\cite{lin04,proakis01,viterbi79}
cover the theory at varying levels of generality but do not provide
spectrum tables, working code, or simulation-validated bound curves
specific to the 802.11 puncture schedules.  Standards documents~\cite{ieee80211}
state the generator polynomials and puncture matrices but say nothing
about performance analysis.  This leaves practitioners either re-deriving
the computation from scratch or relying on unverified figures scattered
across the literature.

This report attempts to fill that gap.  The distance spectrum for each of
the four standard rates is tabulated exactly; union-bound BEP and FER
curves are derived and cross-validated against Monte Carlo simulation for
both QPSK and 256-QAM; and the complete computational procedure is
documented in pseudocode and three reference implementations (Python,
Julia, C++).  We hope the document serves as a useful self-contained
reference for researchers and engineers working with 802.11 physical
layers.

Beyond the 802.11 use case, the method is entirely general: the
augmented-trellis first-return expansion applies to any rate-compatible
punctured convolutional code, regardless of standard or generator
polynomial.  Contributions that broaden the supported code families or improve
the bound computations are welcome.  The Python, Julia, and C++
source code is freely available at
\url{https://github.com/geekymode/bcc\_spectrum}~\cite{bcc_github}.

%% =======================================================
\section{The Mother Code}
%% =======================================================

\subsection{Encoder shift register}

The 802.11 BCC is a \emph{rate-$\frac{1}{2}$, constraint-length-$7$}
convolutional encoder~\cite{forney70} with generator polynomials
\[
  g_1 = \oct{133} = (1,0,1,1,0,1,1),\qquad
  g_2 = \oct{171} = (1,1,1,1,0,0,1).
\]
In polynomial form, with $D$ denoting the unit-delay operator,
\[
  g_1(D) = 1 + D^2 + D^3 + D^5 + D^6,\qquad
  g_2(D) = 1 + D + D^2 + D^3 + D^6.
\]
The two representations are equivalent: the $k$-th coefficient of $g_i(D)$
equals the $k$-th bit of the octal tap mask (LSB = coefficient of $D^0$).
These generators were identified by Odenwalder~\cite{odenwalder70} as
maximising free distance at rate~$\frac{1}{2}$, $K=7$, and were
subsequently standardised for deep-space telemetry by CCSDS~\cite{ccsds}
before being adopted in IEEE~802.11~\cite{ieee80211}.
At each clock cycle the encoder reads one input bit $u_n$, shifts it
into a 6-stage register, and XORs selected taps to produce two output
bits $(v_n^{(1)}, v_n^{(2)})$.

% -----------------------------------------------------------------
% Figure 1 — Encoder
%   Layout: g1 XOR chain ABOVE the delay chain,
%            6 unit-delay ("D") boxes in the middle,
%            g2 XOR chain BELOW the delay chain.
%
%   Position 0 (current input u_n) has NO delay box — just a wire
%   junction (filled dot) at x=0.  The 6 delay boxes D are at
%   x = 1.5, 3.0, 4.5, 6.0, 7.5, 9.0.
%
%   Tap x-coordinates (aligned with junction/box centres):
%     pos 0 = x=0.0  (wire junction, no box)
%     pos 1 = x=1.5  (output of D_1 box)
%     pos 2 = x=3.0  (output of D_2 box)
%     pos 3 = x=4.5  (output of D_3 box)
%     pos 4 = x=6.0  (output of D_4 box, untapped)
%     pos 5 = x=7.5  (output of D_5 box)
%     pos 6 = x=9.0  (output of D_6 box)
%
%   g1 = 133_8: taps at pos 0,2,3,5,6  => XOR nodes at x=0,3,4.5,7.5,9
%   g2 = 171_8: taps at pos 0,1,2,3,6  => XOR nodes at x=0,1.5,3,4.5,9
%   Both chains are on opposite sides => zero crossings by construction.
% -----------------------------------------------------------------
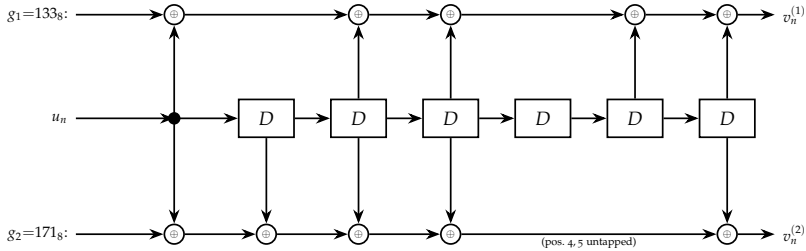
\begin{figure}[ht]
\centering
\resizebox{\linewidth}{!}{%
\begin{tikzpicture}[
  % unit-delay box
  reg/.style  = {draw, rectangle, thick,
                 minimum width=0.90cm, minimum height=0.62cm,
                 font=\small},
  % XOR node
  xorg/.style = {draw, circle, thick,
                 inner sep=0pt, minimum size=0.32cm,
                 font=\tiny},
  >=Stealth, thick
]

%% ---------------------------------------------------------------
%% 6 unit-delay boxes, each labelled "D", at x = 1.5 .. 9.0
%% (position 0 = current input, no box needed)
%% ---------------------------------------------------------------
\node[reg] (Del1) at (1.5, 0) {$D$};
\node[reg] (Del2) at (3.0, 0) {$D$};
\node[reg] (Del3) at (4.5, 0) {$D$};
\node[reg] (Del4) at (6.0, 0) {$D$};
\node[reg] (Del5) at (7.5, 0) {$D$};
\node[reg] (Del6) at (9.0, 0) {$D$};

%% Chain arrows between delay boxes
\foreach \a/\b in {Del1/Del2, Del2/Del3, Del3/Del4, Del4/Del5, Del5/Del6}
  \draw[->] (\a.east) -- (\b.west);

%% Input wire: enters from left, reaches junction TAP0 at x=0,
%% then continues right to first delay box.
\coordinate (TAP0) at (0.0, 0);
\draw[->] (-1.6, 0) -- (TAP0)
  node[pos=0.0, left, font=\scriptsize] {$u_n$};
\draw[->] (TAP0) -- (Del1.west);

%% Filled dot marks the tap-0 wire junction
\filldraw (TAP0) circle (2.5pt);

%% ---------------------------------------------------------------
%% g1 = 133_8 = 1 0 1 1 0 1 1
%% Taps at positions 0,2,3,5,6  =>  x = 0, 3.0, 4.5, 7.5, 9.0
%% XOR chain ABOVE at y = +1.70
%% ---------------------------------------------------------------
\node[xorg] (A0) at (0.0, 1.70) {$\oplus$};
\node[xorg] (A2) at (3.0, 1.70) {$\oplus$};
\node[xorg] (A3) at (4.5, 1.70) {$\oplus$};
\node[xorg] (A5) at (7.5, 1.70) {$\oplus$};
\node[xorg] (A6) at (9.0, 1.70) {$\oplus$};

%% g1 bus
\draw[->] (-1.6, 1.70) -- (A0.west);
\draw[->] (A0.east)    -- (A2.west);
\draw[->] (A2.east)    -- (A3.west);
\draw[->] (A3.east)    -- (A5.west);
\draw[->] (A5.east)    -- (A6.west);
\draw[->] (A6.east)    -- ++(0.6,0)
  node[right, font=\scriptsize] {$v_n^{(1)}$};

%% g1 label
\node[font=\scriptsize, anchor=east] at (-1.6, 1.70)
  {$g_1{=}\oct{133}$:};

%% g1 tap lines (all purely vertical, upward)
%%   tap 0: from wire junction TAP0 straight up to A0
\draw[->] (TAP0) -- (A0.south);
%%   taps 2,3,5,6: from top of respective D box up to XOR node
\draw[->] (Del2.north) -- (A2.south);
\draw[->] (Del3.north) -- (A3.south);
\draw[->] (Del5.north) -- (A5.south);
\draw[->] (Del6.north) -- (A6.south);

%% ---------------------------------------------------------------
%% g2 = 171_8 = 1 1 1 1 0 0 1
%% Taps at positions 0,1,2,3,6  =>  x = 0, 1.5, 3.0, 4.5, 9.0
%% XOR chain BELOW at y = -1.90
%% ---------------------------------------------------------------
\node[xorg] (B0) at (0.0, -1.90) {$\oplus$};
\node[xorg] (B1) at (1.5, -1.90) {$\oplus$};
\node[xorg] (B2) at (3.0, -1.90) {$\oplus$};
\node[xorg] (B3) at (4.5, -1.90) {$\oplus$};
\node[xorg] (B6) at (9.0, -1.90) {$\oplus$};

%% g2 bus
\draw[->] (-1.6, -1.90) -- (B0.west);
\draw[->] (B0.east) -- (B1.west);
\draw[->] (B1.east) -- (B2.west);
\draw[->] (B2.east) -- (B3.west);
\draw[->] (B3.east) -- (B6.west)
  node[midway, below, font=\tiny, inner sep=1pt]
  {(pos.\ 4,\,5 untapped)};
\draw[->] (B6.east) -- ++(0.6,0)
  node[right, font=\scriptsize] {$v_n^{(2)}$};

%% g2 label
\node[font=\scriptsize, anchor=east] at (-1.6, -1.90)
  {$g_2{=}\oct{171}$:};

%% g2 tap lines (all purely vertical, downward)
%%   tap 0: from wire junction TAP0 straight down to B0
\draw[->] (TAP0) -- (B0.north);
%%   taps 1,2,3,6: from bottom of respective D box down to XOR node
\draw[->] (Del1.south) -- (B1.north);
\draw[->] (Del2.south) -- (B2.north);
\draw[->] (Del3.south) -- (B3.north);
\draw[->] (Del6.south) -- (B6.north);

\end{tikzpicture}}% end resizebox
\caption{IEEE~802.11 $K=7$ BCC shift-register encoder.
  Each box labelled $D$ is a unit-delay element; $\oplus$ denotes mod-2
  addition.  The current input $u_n$ enters at the filled dot (tap
  position~0, no delay); the six $D$ elements hold $u_{n-1},\ldots,u_{n-6}$.
  The $g_1=\oct{133}$ XOR chain runs \emph{above} (taps at positions
  0,\,2,\,3,\,5,\,6); the $g_2=\oct{171}$ chain runs \emph{below}
  (taps at positions 0,\,1,\,2,\,3,\,6).
  All tap lines are purely vertical, so no wire crosses any bus segment.}
\label{fig:encoder}
\end{figure}

\subsection{Trellis state}

The \textbf{encoder state} after processing $u_n$ is the 6-bit integer
\[
  \sigma_n = (u_{n-1},u_{n-2},\ldots,u_{n-6}) \in \{0,\ldots,63\}.
\]
A transition $\sigma \xrightarrow{u} \sigma'$ is defined by~\cite{forney70,lin04}
\begin{align}
  \sigma' &= \bigl((\sigma \ll 1) \;\&\; \texttt{0x3F}\bigr) \;\big|\; u,
  \label{eq:state-update}\\
  v^{(k)} &= \bigoplus_{i=0}^{6} g_k[i]\cdot r_i, \quad
  \mathbf{r} = (u,\sigma_0,\ldots,\sigma_5). \label{eq:output}
\end{align}
The resulting trellis~\cite{forney73} has $2^{K-1}=64$ states, each with exactly
two outgoing branches (for $u\in\{0,1\}$).

\begin{figure*}[ht]
\centering
\begin{tikzpicture}[
  snode/.style = {circle, draw, thick,
                  minimum size=0.28cm, font=\tiny, inner sep=0pt},
  e0/.style    = {->, >=Stealth, thick, draw=tufteblue},
  e1/.style    = {->, >=Stealth, thick, draw=tuftered, dashed},
  lbl/.style   = {font=\tiny, fill=white, inner sep=1pt},
  xstep=2.1, ystep=1.15
]

% State nodes: 4 states x 8 time columns
\foreach \t in {0,1,...,7}{
  \foreach \s in {0,1,2,3}{
    \node[snode] (N\t\s) at (\t*2.1, -\s*1.15) {};
  }
  \node[above, font=\scriptsize] at (\t*2.1, 0.42) {$n{=}\t$};
}

% State labels left (n=0) and right (n=7)
\foreach \s/\lbl in {0/00, 1/01, 2/10, 3/11}{
  \node[left=0.18cm  of N0\s, font=\scriptsize] {$\sigma{=}\lbl$};
  \node[right=0.18cm of N7\s, font=\scriptsize] {$\lbl$};
}

% Transitions n=0->1 WITH output labels
% State 00: u=0->00 out=00, u=1->10 out=11
\draw[e0] (N00) to node[lbl,above,sloped]{00} (N10);
\draw[e1] (N00) to node[lbl,below,sloped]{11} (N12);
% State 01: u=0->00 out=11, u=1->10 out=00
\draw[e0] (N01) to node[lbl,above,sloped]{11} (N10);
\draw[e1] (N01) to node[lbl,below,sloped]{00} (N12);
% State 10: u=0->01 out=10, u=1->11 out=01
\draw[e0] (N02) to node[lbl,above,sloped]{10} (N11);
\draw[e1] (N02) to node[lbl,below,sloped]{01} (N13);
% State 11: u=0->01 out=01, u=1->11 out=10
\draw[e0] (N03) to node[lbl,above,sloped]{01} (N11);
\draw[e1] (N03) to node[lbl,below,sloped]{10} (N13);

% Remaining transitions without labels
\foreach \t in {1,2,...,6}{
  \pgfmathtruncatemacro{\tn}{\t+1}
  \draw[e0] (N\t0) -- (N\tn0);  \draw[e1] (N\t0) -- (N\tn2);
  \draw[e0] (N\t1) -- (N\tn0);  \draw[e1] (N\t1) -- (N\tn2);
  \draw[e0] (N\t2) -- (N\tn1);  \draw[e1] (N\t2) -- (N\tn3);
  \draw[e0] (N\t3) -- (N\tn1);  \draw[e1] (N\t3) -- (N\tn3);
}

% Inline legend
\node[font=\scriptsize, anchor=west] at (0.0, -5.0) {%
  \textcolor{tufteblue}{---}\;$u=0$ (blue solid)\quad
  \textcolor{tuftered}{- -}\;$u=1$ (red dashed)\quad
  Edge labels at $n{=}0$: $(v^{(1)}v^{(2)})$
};

\end{tikzpicture}
\caption{Rate-$\frac{1}{2}$ trellis, 4 lowest states over $n=0,\ldots,7$.
  Blue solid: $u=0$; red dashed: $u=1$.
  Edge labels $(v^{(1)}v^{(2)})$ are shown for the first transition only;
  the pattern repeats identically at every subsequent step.
  The full code has 64 states; here $\sigma\in\{00,01,10,11\}_2$.}
\label{fig:trellis}
\end{figure*}
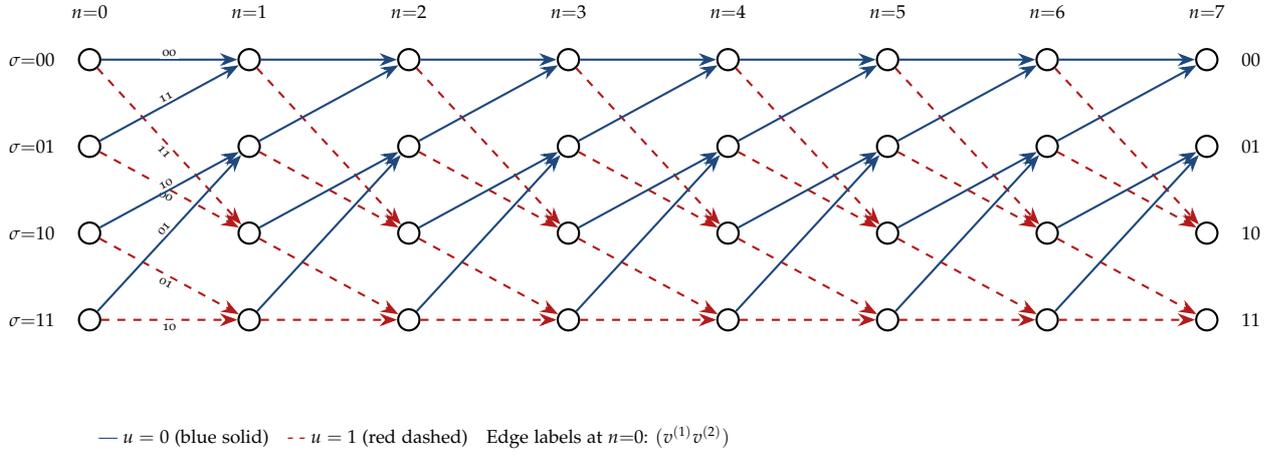

Figure~\ref{fig:trellis} shows the first four states over seven time steps.

% -----------------------------------------------------------------
% Figure 2 — Trellis (full-width figure*)
%   * 4 states, n=0..7 (8 columns).
%   * No shading; no legend boxes — inline note only.
%   * Edge labels shown for n=0->1 only; same pattern repeats.
% -----------------------------------------------------------------

%% =======================================================
\section{Puncturing and Puncture Matrices}
%% =======================================================

Higher code rates are obtained by \emph{puncturing}: periodically deleting
output bits before transmission~\cite{cain79,yasuda84,hagenauer88}.
Cain et~al.~\cite{cain79} showed that rate-$(n{-}1)/n$ codes can be derived
this way from a low-rate mother code without re-designing the decoder;
Yasuda et~al.~\cite{yasuda84} extended the approach to high rates with
soft-decision Viterbi decoding; and Hagenauer~\cite{hagenauer88} unified the
family into rate-compatible punctured convolutional (RCPC) codes.
The surviving bits are specified by a binary \emph{puncture matrix}
$\mathbf{P}$, whose two rows correspond to the output streams $v^{(1)}$
and $v^{(2)}$, and whose columns span one period.
A~$1$ means ``transmit''; a~$0$ means ``delete''.
The serialised puncture mask $\mathbf{p}$ is read column-by-column from
$\mathbf{P}$.

\subsection{Rate \nicefrac{1}{2} — unpunctured}

Both outputs are kept at every step.  The period is $L=2$:
\[
  \mathbf{P}_{1/2} =
  \begin{pmatrix}1\\1\end{pmatrix},
  \qquad \mathbf{p} = (1,1).
\]

\subsection{Rate \nicefrac{2}{3}}

For every two input bits the encoder produces 4 output bits; 3 are
transmitted ($v^{(2)}_1$ is deleted):
\[
  \mathbf{P}_{2/3} =
  \begin{pmatrix}1 & 1\\ 1 & 0\end{pmatrix},
  \qquad \mathbf{p} = (1,1,1,0),\quad L=4.
\]
Code rate $= 2/3$ because 2 info bits yield 3 transmitted bits.

\subsection{Rate \nicefrac{3}{4}}

Three input bits produce 6 outputs; 4 are transmitted:
\[
  \mathbf{P}_{3/4} =
  \begin{pmatrix}1 & 1 & 0\\ 1 & 0 & 1\end{pmatrix},
  \qquad \mathbf{p} = (1,1,1,0,0,1),\quad L=6.
\]
Positions 4 and 5 of the serial stream ($v^{(1)}_2$ and $v^{(2)}_1$)
are deleted.

\subsection{Rate \nicefrac{5}{6}}

Five input bits produce 10 outputs; 6 are transmitted:
\[
  \mathbf{P}_{5/6} =
  \begin{pmatrix}1 & 1 & 0 & 1 & 0\\ 1 & 0 & 1 & 0 & 1\end{pmatrix},
  \qquad \mathbf{p} = (1,1,1,0,0,1,1,0,1,0),\quad L=10.
\]

\begin{margintable}
\centering\small
\begin{tabular}{@{}llcc@{}}
\toprule
Rate & Mask $\mathbf{p}$ & $L$ & $d_\mathrm{free}$\\
\midrule
$\frac{1}{2}$ & \texttt{11}         &  2 & 10\\
$\frac{2}{3}$ & \texttt{1110}       &  4 &  6\\
$\frac{3}{4}$ & \texttt{111001}     &  6 &  5\\
$\frac{5}{6}$ & \texttt{1110011010} & 10 &  4\\
\bottomrule
\end{tabular}
\caption{802.11 puncture masks and free distances.}
\label{tab:masks}
\end{margintable}

Figure~\ref{fig:puncture} illustrates the rate-\nicefrac{3}{4} mask over
two full periods.  Solid-bordered cells are transmitted; dashed cells are
deleted before the channel.

% -----------------------------------------------------------------
% Figure 3 — Puncture illustration
%   * No shading. No legend boxes.
%   * Labels explicit (no pgfmath array indexing).
%   * Transmitted = solid border; deleted = dashed border + grey text.
% -----------------------------------------------------------------
\begin{figure}[ht]
\centering
\begin{tikzpicture}[
  kept/.style = {draw, thick, rectangle,
                 minimum width=0.82cm, minimum height=0.54cm,
                 font=\small},
  gone/.style = {draw, thick, dashed, rectangle,
                 minimum width=0.82cm, minimum height=0.54cm,
                 font=\small, text=gray!70},
  >=Stealth
]

%% Period 1:  mask = 1 1 1 0 0 1
\node[kept] (c0)  at ( 0*0.90, 0) {$v_0^1$};
\node[kept] (c1)  at ( 1*0.90, 0) {$v_0^2$};
\node[kept] (c2)  at ( 2*0.90, 0) {$v_1^1$};
\node[gone] (c3)  at ( 3*0.90, 0) {$v_1^2$};
\node[gone] (c4)  at ( 4*0.90, 0) {$v_2^1$};
\node[kept] (c5)  at ( 5*0.90, 0) {$v_2^2$};

%% Period 2:  same mask repeated
\node[kept] (c6)  at ( 6*0.90, 0) {$v_3^1$};
\node[kept] (c7)  at ( 7*0.90, 0) {$v_3^2$};
\node[kept] (c8)  at ( 8*0.90, 0) {$v_4^1$};
\node[gone] (c9)  at ( 9*0.90, 0) {$v_4^2$};
\node[gone] (c10) at (10*0.90, 0) {$v_5^1$};
\node[kept] (c11) at (11*0.90, 0) {$v_5^2$};

%% Braces
\draw[decorate,decoration={brace,amplitude=5pt,mirror}]
  (-0.41,-0.40) -- (5*0.90+0.41,-0.40)
  node[midway,below=6pt,font=\scriptsize]{period 1 ($L=6$)};
\draw[decorate,decoration={brace,amplitude=5pt,mirror}]
  (6*0.90-0.41,-0.40) -- (11*0.90+0.41,-0.40)
  node[midway,below=6pt,font=\scriptsize]{period 2};

\end{tikzpicture}
\caption{Rate-$\frac{3}{4}$ mask \texttt{111001} over two periods.
  Solid border: transmitted.  Dashed border: deleted before transmission.
  Superscript indexes the generator ($1$ or $2$); subscript indexes
  the input-bit time step.
  4 of every 6 serial output bits are transmitted, giving rate $\frac{4/2}{6/2}=\frac{2}{3}\cdot\frac{3}{2}=\frac{3}{4}$.}
\label{fig:puncture}
\end{figure}

%% =======================================================
\section{Distance Spectrum: Definitions}
%% =======================================================

\begin{definition}[Distance spectrum \cite{lin04,ryan09,viterbi79}]
The \emph{distance spectrum} of a convolutional code is the pair of
integer sequences $\{(\alpha_d,\beta_d)\}_{d\geq d_\mathrm{free}}$ where
$\alpha_d$ counts error events at Hamming distance~$d$ from the all-zero
codeword, and $\beta_d$ is the total input Hamming weight of those paths.
The smallest $d$ for which $\alpha_d>0$ is the \emph{free distance}
$d_\mathrm{free}$~\cite{viterbi67,forney70}, the primary figure of merit
for a convolutional code's error-correcting capability.
\end{definition}

%% =======================================================
\section{BER Bounds from the Distance Spectrum}
%% =======================================================

\subsection{Union bound on bit-error probability}

For the Viterbi decoder~\cite{viterbi67,forney73} over an AWGN channel
with BPSK modulation, the pairwise error probability between two codewords
at Hamming distance~$d$ is~\cite{proakis01,ryan09}
\[
  P_2(d) = Q\!\left(\sqrt{2d\,R_c\,\frac{E_b}{N_0}}\right),
\]
where $R_c$ is the code rate, $E_b/N_0$ is the information-bit SNR, and
$Q(x)=\frac{1}{\sqrt{2\pi}}\int_x^\infty e^{-t^2/2}\,\mathrm{d}t$.
Applying the union bound over all error events~\cite{viterbi67,lin04}
gives the \emph{union-bound bit-error probability}
\begin{equation}
  P_b \;\lesssim\;
  \sum_{d=d_\mathrm{free}}^{D_{\max}}
    \beta_d \; Q\!\left(\sqrt{2d\,R_c\,\frac{E_b}{N_0}}\right),
  \label{eq:ub}
\end{equation}
where $k=1$ information bit per step is used (absorbed into $\beta_d$).

\subsection{Frame error rate}

The \emph{frame/event error rate} uses the multiplicity directly:
\begin{equation}
  P_f \;\lesssim\;
  \sum_{d=d_\mathrm{free}}^{D_{\max}}
    \alpha_d \; Q\!\left(\sqrt{2d\,R_c\,\frac{E_b}{N_0}}\right).
  \label{eq:fer}
\end{equation}

\subsection{Single-term approximation}

Near the waterfall region the leading term dominates:
\begin{equation}
  P_b \;\approx\;
  \beta_{d_\mathrm{free}} \;
  Q\!\left(\sqrt{2\,d_\mathrm{free}\,R_c\,\frac{E_b}{N_0}}\right).
  \label{eq:approx}
\end{equation}
Table~\ref{tab:ber_table} evaluates the full bound~\eqref{eq:ub} using
10~spectrum terms at several SNR points.

\begin{table*}[ht]
\centering
\small
\setlength{\tabcolsep}{5.5pt}
\begin{tabular}{@{}l cc cc cc cc@{}}
\toprule
 & \multicolumn{2}{c}{\textbf{Rate $\frac{1}{2}$}}
 & \multicolumn{2}{c}{\textbf{Rate $\frac{2}{3}$}}
 & \multicolumn{2}{c}{\textbf{Rate $\frac{3}{4}$}}
 & \multicolumn{2}{c}{\textbf{Rate $\frac{5}{6}$}} \\
\cmidrule(lr){2-3}\cmidrule(lr){4-5}\cmidrule(lr){6-7}\cmidrule(lr){8-9}
$E_b/N_0$ (dB)
 & $P_b^{\rm UB}$ & $P_f^{\rm UB}$
 & $P_b^{\rm UB}$ & $P_f^{\rm UB}$
 & $P_b^{\rm UB}$ & $P_f^{\rm UB}$
 & $P_b^{\rm UB}$ & $P_f^{\rm UB}$ \\
\midrule
3 & $1.6\!\times\!10^{-1}$ & $1.8\!\times\!10^{-1}$
  & $1.4\!\times\!10^{-1}$ & $1.4\!\times\!10^{-1}$
  & $1.3\!\times\!10^{-1}$ & $1.1\!\times\!10^{-1}$
  & $1.2\!\times\!10^{-1}$ & $1.7\!\times\!10^{-1}$\\
4 & $4.2\!\times\!10^{-2}$ & $5.3\!\times\!10^{-2}$
  & $5.1\!\times\!10^{-2}$ & $5.3\!\times\!10^{-2}$
  & $6.3\!\times\!10^{-2}$ & $5.5\!\times\!10^{-2}$
  & $8.1\!\times\!10^{-2}$ & $1.1\!\times\!10^{-1}$\\
5 & $4.2\!\times\!10^{-4}$ & $5.3\!\times\!10^{-4}$
  & $8.7\!\times\!10^{-3}$ & $9.0\!\times\!10^{-3}$
  & $1.7\!\times\!10^{-2}$ & $1.5\!\times\!10^{-2}$
  & $4.4\!\times\!10^{-2}$ & $6.2\!\times\!10^{-2}$\\
6 & $5.8\!\times\!10^{-8}$ & $7.2\!\times\!10^{-8}$
  & $4.4\!\times\!10^{-4}$ & $4.5\!\times\!10^{-4}$
  & $2.0\!\times\!10^{-3}$ & $1.7\!\times\!10^{-3}$
  & $1.3\!\times\!10^{-2}$ & $1.8\!\times\!10^{-2}$\\
7 & $<10^{-14}$            & $<10^{-14}$
  & $3.6\!\times\!10^{-6}$ & $3.7\!\times\!10^{-6}$
  & $5.4\!\times\!10^{-5}$ & $4.7\!\times\!10^{-5}$
  & $1.9\!\times\!10^{-3}$ & $2.6\!\times\!10^{-3}$\\
8 & $\ll10^{-20}$          & $\ll10^{-20}$
  & $3.5\!\times\!10^{-9}$ & $3.6\!\times\!10^{-9}$
  & $2.7\!\times\!10^{-7}$ & $2.4\!\times\!10^{-7}$
  & $1.4\!\times\!10^{-4}$ & $2.0\!\times\!10^{-4}$\\
\bottomrule
\end{tabular}
\caption{Union-bound BEP~\eqref{eq:ub} and FER~\eqref{eq:fer} using
  10~spectrum terms, BPSK over AWGN~\protect\cite{proakis01,viterbi67}.
  The rate-$\frac{1}{2}$ bound collapses rapidly above 5\,dB;
  higher-rate codes require higher SNR for the same error floor because
  their spectra are denser and $d_\mathrm{free}$ is smaller.}
\label{tab:ber_table}
\end{table*}

\subsection{Notes on bound tightness}

The union bound~\eqref{eq:ub} is loose at low SNR (overlapping pairwise
events) but tightens rapidly in the waterfall region because
$Q(\sqrt{2d_\mathrm{free}R_c E_b/N_0})$ decays faster than any polynomial~\cite{forney73}.
For the rate-$\frac{1}{2}$ code the single $d=10$ term~\eqref{eq:approx}
is already tight above $5\,\mathrm{dB}$.  Higher-rate codes require several
more terms because their spectra grow more quickly with~$d$.

\begin{figure*}[ht]
\centering
\begin{tikzpicture}
\begin{groupplot}[
  group style={
    group size=2 by 1,
    horizontal sep=2.4cm,
  },
  width=0.46\textwidth, height=0.46\textwidth,
  xmin=-2, xmax=10,
  ymode=log, ymin=1e-9, ymax=1.5,
  xlabel={$E_b/N_0$ (dB)},
  grid=both,
  grid style={line width=0.3pt, draw=gray!30},
  major grid style={line width=0.5pt, draw=gray!50},
  tick label style={font=\scriptsize},
  label style={font=\small},
  legend style={font=\scriptsize, at={(0.03,0.03)},
    anchor=south west, draw=gray!60, fill=white,
    inner sep=3pt, row sep=-1pt},
  legend cell align=left,
  clip=true,
]
%% ============================================================
%% LEFT PANEL — BEP upper bound P_b  +  simulation markers
%% ============================================================
\nextgroupplot[ylabel={BER $P_b$}]

%% Uncoded BPSK reference: Q(sqrt(2*Eb/N0))
\addplot[black, densely dotted, thick] coordinates {
  (-2.00,1.3537e-01) (-1.00,1.0567e-01) (0.00,7.8650e-02) (0.25,7.2764e-02)
  (0.50,6.7065e-02) (0.75,6.1567e-02) (1.00,5.6282e-02) (1.25,5.1223e-02)
  (1.50,4.6401e-02) (1.75,4.1826e-02) (2.00,3.7506e-02) (2.25,3.3448e-02)
  (2.50,2.9655e-02) (2.75,2.6132e-02) (3.00,2.2878e-02) (3.25,1.9893e-02)
  (3.50,1.7173e-02) (3.75,1.4711e-02) (4.00,1.2501e-02) (4.25,1.0532e-02)
  (4.50,8.7938e-03) (4.75,7.2725e-03) (5.00,5.9539e-03) (5.25,4.8225e-03)
  (5.50,3.8622e-03) (5.75,3.0564e-03) (6.00,2.3883e-03) (6.25,1.8414e-03)
  (6.50,1.3998e-03) (6.75,1.0483e-03) (7.00,7.7267e-04) (7.25,5.6005e-04)
  (7.50,3.9880e-04) (7.75,2.7868e-04) (8.00,1.9091e-04) (8.25,1.2805e-04)
  (8.50,8.4000e-05) (8.75,5.3816e-05) (9.00,3.3627e-05) (9.25,2.0464e-05)
  (9.50,1.2109e-05) (9.75,6.9558e-06) (10.00,3.8721e-06)
};
\addlegendentry{Uncoded BPSK}

%% ---- Rate 1/2: bound (solid) + sim markers ----
\addplot[tufteblue, solid, thick] coordinates {
  (2.25,2.053e-01) (2.50,1.993e-02) (2.75,3.333e-03) (3.00,8.896e-04)
  (3.25,3.033e-04) (3.50,1.155e-04) (3.75,4.619e-05) (4.00,1.878e-05)
  (4.25,7.608e-06) (4.50,3.033e-06) (4.75,1.178e-06) (5.00,4.427e-07)
  (5.25,1.598e-07) (5.50,5.514e-08) (5.75,1.808e-08) (6.00,5.609e-09)
  (6.25,1.638e-09) (6.50,4.480e-10) (6.75,1.143e-10) (7.00,2.702e-11)
};
\addlegendentry{$R=1/2$ (bound)}
\addplot[tufteblue, dashdotted, thick, mark=*, mark size=2pt] coordinates {
  (0.00,1.5011e-01) (1.00,3.8892e-02) (2.00,4.8939e-03) (3.00,3.5851e-04)
  (4.00,1.5317e-05) (5.00,3.3368e-07) (6.00,6.5226e-09)
};
\addlegendentry{$R=1/2$ (sim)}

%% ---- Rate 2/3: bound (solid) + sim markers ----
\addplot[tuftered, solid, thick] coordinates {
  (2.75,3.093e-01) (3.00,4.374e-02) (3.25,7.863e-03) (3.50,1.914e-03)
  (3.75,5.904e-04) (4.00,2.084e-04) (4.25,7.856e-05) (4.50,3.043e-05)
  (4.75,1.184e-05) (5.00,4.567e-06) (5.25,1.727e-06) (5.50,6.361e-07)
  (5.75,2.268e-07) (6.00,7.786e-08) (6.25,2.564e-08) (6.50,8.065e-09)
  (6.75,2.414e-09) (7.00,6.855e-10) (7.25,1.839e-10) (7.50,4.644e-11)
};
\addlegendentry{$R=2/3$ (bound)}
\addplot[tuftered, dashdotted, thick, mark=square*, mark size=2pt] coordinates {
  (0.00,2.9075e-01) (1.00,1.2032e-01) (2.00,2.1314e-02) (3.00,1.5709e-03)
  (4.00,7.1476e-05) (5.00,2.0767e-06) (6.00,4.1114e-08)
};
\addlegendentry{$R=2/3$ (sim)}

%% ---- Rate 3/4: bound (solid) + sim markers ----
\addplot[tuftegreen, solid, thick] coordinates {
  (3.25,3.834e-01) (3.50,4.226e-02) (3.75,7.508e-03) (4.00,2.050e-03)
  (4.25,7.111e-04) (4.50,2.763e-04) (4.75,1.132e-04) (5.00,4.732e-05)
  (5.25,1.981e-05) (5.50,8.192e-06) (5.75,3.316e-06) (6.00,1.303e-06)
  (6.25,4.942e-07) (6.50,1.798e-07) (6.75,6.240e-08) (7.00,2.057e-08)
  (7.25,6.408e-09) (7.50,1.878e-09) (7.75,5.152e-10) (8.00,1.317e-10)
};
\addlegendentry{$R=3/4$ (bound)}
\addplot[tuftegreen, dashdotted, thick, mark=triangle*, mark size=2.5pt] coordinates {
  (-2.00,4.7563e-01) (-1.00,4.5034e-01) (0.00,3.7064e-01) (1.00,2.2987e-01)
  (2.00,5.9396e-02) (3.00,5.4355e-03) (4.00,3.1007e-04)
  (5.00,1.2662e-05) (6.00,4.1961e-07) (7.00,6.6503e-09)
};
\addlegendentry{$R=3/4$ (sim)}

%% ---- Rate 5/6: bound (solid) + sim markers ----
\addplot[violet, solid, thick] coordinates {
  (4.00,1.191e-01) (4.25,1.519e-02) (4.50,3.750e-03) (4.75,1.273e-03)
  (5.00,4.964e-04) (5.25,2.062e-04) (5.50,8.782e-05) (5.75,3.754e-05)
  (6.00,1.587e-05) (6.25,6.564e-06) (6.50,2.634e-06) (6.75,1.019e-06)
  (7.00,3.773e-07) (7.25,1.332e-07) (7.50,4.459e-08) (7.75,1.409e-08)
  (8.00,4.182e-09) (8.25,1.161e-09) (8.50,2.998e-10) (8.75,7.171e-11)
};
\addlegendentry{$R=5/6$ (bound)}
\addplot[violet, dashdotted, thick, mark=diamond*, mark size=2.5pt] coordinates {
  (-1.00,4.6819e-01) (0.00,4.3327e-01) (1.00,3.4382e-01) (2.00,1.6398e-01)
  (3.00,2.6043e-02) (4.00,1.8482e-03) (5.00,9.3410e-05)
  (6.00,2.6955e-06) (7.00,6.1649e-08)
};
\addlegendentry{$R=5/6$ (sim)}

%% ============================================================
%% RIGHT PANEL — FER upper bound P_f  +  simulation markers
%% ============================================================
\nextgroupplot[ylabel={FER $P_f$}]

%% ---- Rate 1/2: bound (dashed) + sim markers ----
\addplot[tufteblue, solid, thick] coordinates {
  (2.75,3.0237e-01) (3.00,1.1706e-01) (3.25,5.0261e-02) (3.50,2.2331e-02)
  (3.75,9.9615e-03) (4.00,4.3897e-03) (4.25,1.8912e-03) (4.50,7.9065e-04)
  (4.75,3.1884e-04) (5.00,1.2340e-04) (5.25,4.5622e-05) (5.50,1.6043e-05)
  (5.75,5.3431e-06) (6.00,1.6783e-06) (6.25,4.9501e-07) (6.50,1.3651e-07)
  (6.75,3.5033e-08) (7.00,8.3279e-09) (7.25,1.8246e-09)
};
\addlegendentry{$R=1/2$ (bound)}
\addplot[tufteblue, dashdotted, thick, mark=*, mark size=2pt] coordinates {
  (1.00,9.7754e-01) (2.00,4.8633e-01) (3.00,6.3151e-02) (4.00,3.7007e-03)
  (5.00,9.6016e-05) (6.00,1.8851e-06)
};
\addlegendentry{$R=1/2$ (sim)}

%% ---- Rate 2/3: bound (dashed) + sim markers ----
\addplot[tuftered, solid, thick] coordinates {
  (3.25,5.1349e-01) (3.50,1.8034e-01) (3.75,7.2219e-02) (4.00,3.0521e-02)
  (4.25,1.3104e-02) (4.50,5.6033e-03) (4.75,2.3587e-03) (5.00,9.6974e-04)
  (5.25,3.8713e-04) (5.50,1.4936e-04) (5.75,5.5471e-05) (6.00,1.9761e-05)
  (6.25,6.7295e-06) (6.50,2.1837e-06) (6.75,6.7299e-07) (7.00,1.9631e-07)
  (7.25,5.4008e-08) (7.50,1.3959e-08) (7.75,3.3754e-09)
};
\addlegendentry{$R=2/3$ (bound)}
\addplot[tuftered, dashdotted, thick, mark=square*, mark size=2pt] coordinates {
  (1.00,9.9935e-01) (2.00,8.0078e-01) (3.00,1.6155e-01) (4.00,1.1636e-02)
  (5.00,4.8472e-04) (6.00,9.8595e-06) (7.00,1.3000e-07)
};
\addlegendentry{$R=2/3$ (sim)}

%% ---- Rate 3/4: bound (dashed) + sim markers ----
\addplot[tuftegreen, solid, thick] coordinates {
  (3.75,4.7737e-01) (4.00,1.8565e-01) (4.25,8.0323e-02) (4.50,3.6124e-02)
  (4.75,1.6385e-02) (5.00,7.3748e-03) (5.25,3.2596e-03) (5.50,1.4042e-03)
  (5.75,5.8603e-04) (6.00,2.3575e-04) (6.25,9.0996e-05) (6.50,3.3553e-05)
  (6.75,1.1769e-05) (7.00,3.9109e-06) (7.25,1.2260e-06) (7.50,3.6099e-07)
  (7.75,9.9402e-08) (8.00,2.5479e-08) (8.25,6.0504e-09) (8.50,1.3245e-09)
};
\addlegendentry{$R=3/4$ (bound)}
\addplot[tuftegreen, dashdotted, thick, mark=triangle*, mark size=2.5pt] coordinates {
  (2.00,9.4340e-01) (3.00,3.4843e-01) (4.00,3.6350e-02) (5.00,1.9785e-03)
  (6.00,7.2471e-05) (7.00,1.2384e-06)
};
\addlegendentry{$R=3/4$ (sim)}

%% ---- Rate 5/6: bound (dashed) + sim markers ----
\addplot[violet, solid, thick] coordinates {
  (4.25,7.2526e-01) (4.50,2.7597e-01) (4.75,1.2016e-01) (5.00,5.4777e-02)
  (5.25,2.5246e-02) (5.50,1.1555e-02) (5.75,5.1933e-03) (6.00,2.2741e-03)
  (6.25,9.6418e-04) (6.50,3.9376e-04) (6.75,1.5416e-04) (7.00,5.7612e-05)
  (7.25,2.0464e-05) (7.50,6.8803e-06) (7.75,2.1804e-06) (8.00,6.4852e-07)
  (8.25,1.8025e-07) (8.50,4.6604e-08) (8.75,1.1156e-08) (9.00,2.4606e-09)
};
\addlegendentry{$R=5/6$ (bound)}
\addplot[violet, dashdotted, thick, mark=diamond*, mark size=2.5pt] coordinates {
  (2.00,9.9642e-01) (3.00,7.1191e-01) (4.00,1.3086e-01) (5.00,1.0954e-02)
  (6.00,4.0095e-04) (7.00,9.4844e-06) (8.00,1.3000e-07)
};
\addlegendentry{$R=5/6$ (sim)}

\end{groupplot}
\end{tikzpicture}
\caption{BEP union bound~\eqref{eq:ub} (left) and FER union bound~\eqref{eq:fer}
  (right, $K=1024$ information bits) for all four IEEE~802.11 BCC rates over AWGN
  with QPSK modulation~\cite{proakis01,viterbi67}.
  Lines: union bounds using 30~spectrum terms ($d_{\max}=130$), solid for both BEP and FER.
  Markers: Monte Carlo simulation, dot-dashed lines;
  $\bullet$\,$R=\frac{1}{2}$,
  $\blacksquare$\,$R=\frac{2}{3}$,
  $\blacktriangle$\,$R=\frac{3}{4}$,
  $\blacklozenge$\,$R=\frac{5}{6}$.
  Dotted black: uncoded BPSK $Q\!\left(\sqrt{2E_b/N_0}\right)$ (BEP panel only).
  The FER bound lies roughly $K/\bar{w}$ times above the BEP bound
  ($\bar{w}\approx3$--$7$ per rate); both bounds track the simulation
  closely in the waterfall region.}
\label{fig:ber_fer}
\end{figure*}
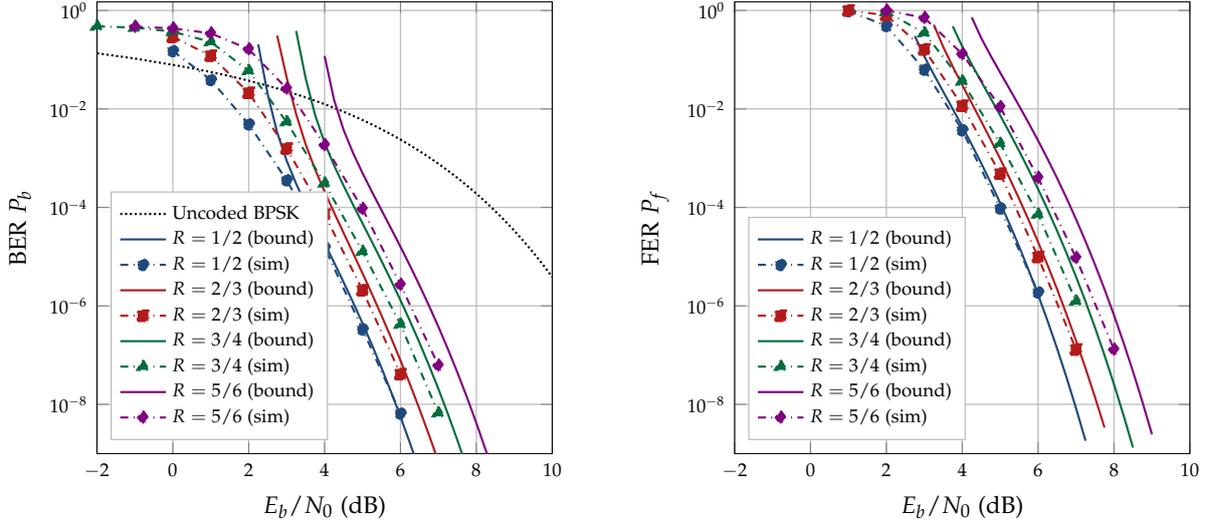

Uncoded BPSK has $P_b = Q(\sqrt{2E_b/N_0})$~\cite{proakis01}.
The rate-$\frac{1}{2}$ code provides roughly 4--5\,dB coding gain at
$P_b=10^{-5}$~\cite{heller71,proakis01}; the rate-$\frac{3}{4}$ code
provides approximately 2--3\,dB, trading protection for bandwidth efficiency.
Figure~\ref{fig:ber_fer} plots both bounds across the full waterfall region.

% -----------------------------------------------------------------
% Figure: BER and FER union-bound curves + simulation markers
%   Left panel:  BEP  P_b  (eq:ub)  + sim markers (filled circles)
%   Right panel: FER  P_f  (eq:fer) + sim markers (filled squares)
%   Simulation: QPSK/AWGN, all four rates.
%   Uncoded BPSK shown as black dotted reference.
% -----------------------------------------------------------------

%% =======================================================
\section{Bounds for Gray-Coded \textit{M}-QAM}
%% =======================================================

\subsection{Channel model}

In IEEE~802.11 the BCC-encoded bit stream is interleaved and mapped to
$M$-QAM symbols.  Under the standard AWGN channel model, the received
complex sample for symbol index $k$ is
\begin{equation}
  y_k = \sqrt{E_s}\,s_k + n_k, \qquad n_k \sim \mathcal{CN}(0,N_0),
\end{equation}
where $s_k\in\mathcal{S}$ is the transmitted symbol from the normalised
Gray-coded square $M$-QAM alphabet~\cite{proakis01,ryan09}, and
$E_s = m\,R_c\,E_b$ with $m=\log_2 M$ bits per symbol.

\subsection{Effective SNR factor $\Delta_M$}

A Gray-coded soft demapper computes one LLR per coded bit.  Under
the BICM model~\cite{caire98}, the dominant contribution to the pairwise
error probability between paths at Hamming distance~$d$ uses the
minimum intra-constellation bit distance, giving the upper bound~\cite{proakis01,lin04}
\begin{equation}
  P_2(d) \leq Q\!\left(\sqrt{2\,R_c\,d\,\Delta_M\,\frac{E_b}{N_0}}\right),
  \label{eq:pep_qam}
\end{equation}
where the \emph{modulation penalty factor}
\begin{equation}
  \Delta_M = \frac{3\,m}{2(M-1)}, \qquad m = \log_2 M,\quad M\geq 4,
  \label{eq:deltaM}
\end{equation}
captures the reduced per-bit minimum Euclidean distance as the constellation
grows.  For QPSK ($M=4$) the formula yields $\Delta_M=1$, recovering the
standard BPSK result; BPSK ($M=2$) is the underlying binary channel
to which \eqref{eq:deltaM} does not apply.
Table~\ref{tab:deltaM} lists the values for the four constellations used
in IEEE~802.11~\cite{ieee80211}.

\begin{table}[ht]
\centering\small
\begin{tabular}{@{} c c c r @{}}
\toprule
$M$ & $m$ & $\Delta_M$ & Penalty (dB) \\
\midrule
4   & 2 & $1$                        &  $0.0$ \\
16  & 4 & $\nicefrac{2}{5}=0.4$      & $-4.0$ \\
64  & 6 & $\nicefrac{2}{14}\approx0.143$ & $-8.5$ \\
256 & 8 & $\nicefrac{8}{170}\approx0.047$ & $-13.3$ \\
\bottomrule
\end{tabular}
\caption{Modulation penalty factor~$\Delta_M$~\eqref{eq:deltaM} and
  equivalent $E_b/N_0$ penalty relative to BPSK at the same coded-bit
  reliability.  Each step from QPSK to the next order costs roughly
  4--5\,dB.}
\label{tab:deltaM}
\end{table}

\subsection{Union bounds}

Substituting \eqref{eq:pep_qam} into the BPSK bounds replaces
$E_b/N_0$ by $\Delta_M\,E_b/N_0$, giving the $M$-QAM BEP and FER
union bounds~\cite{proakis01,lin04}:
\begin{align}
  P_b &\leq \sum_{d=d_\mathrm{free}}^{\infty}
          \beta_d\,Q\!\left(\sqrt{2\,R_c\,d\,\Delta_M\,\frac{E_b}{N_0}}\right),
  \label{eq:ub_qam}\\
  P_f(K) &\leq K\!\sum_{d=d_\mathrm{free}}^{\infty}
          \alpha_d\,Q\!\left(\sqrt{2\,R_c\,d\,\Delta_M\,\frac{E_b}{N_0}}\right).
  \label{eq:fer_qam}
\end{align}
The factor $K$ is the frame length in information bits; it enters because
any of the $K$ trellis steps can initiate an error event.
Setting $\Delta_M=1$ (QPSK) recovers the BPSK bounds~\eqref{eq:ub}
and~\eqref{eq:fer} exactly.  Higher-order QAM shifts the entire waterfall
to the right by $10\log_{10}(1/\Delta_M)$\,dB with no change in slope.

The uncoded reference BEP for Gray-coded $M$-QAM is
\begin{equation}
  P_b^{\rm unc}(M) = \frac{4}{m}\!\left(1-\frac{1}{\sqrt{M}}\right)
    Q\!\left(\sqrt{\frac{3\,m}{M-1}\,\frac{E_b}{N_0}}\right),
\end{equation}
a tight approximation for $E_b/N_0 \gtrsim 5$\,dB~\cite{proakis01}.

% -----------------------------------------------------------------
% Figure: BEP union bounds for M-QAM
%   Left  panel: Rate 1/2, M in {4,16,64,256}
%   Right panel: 16-QAM,   all four code rates
%   Uncoded M-QAM shown as dotted reference in each panel.
% -----------------------------------------------------------------
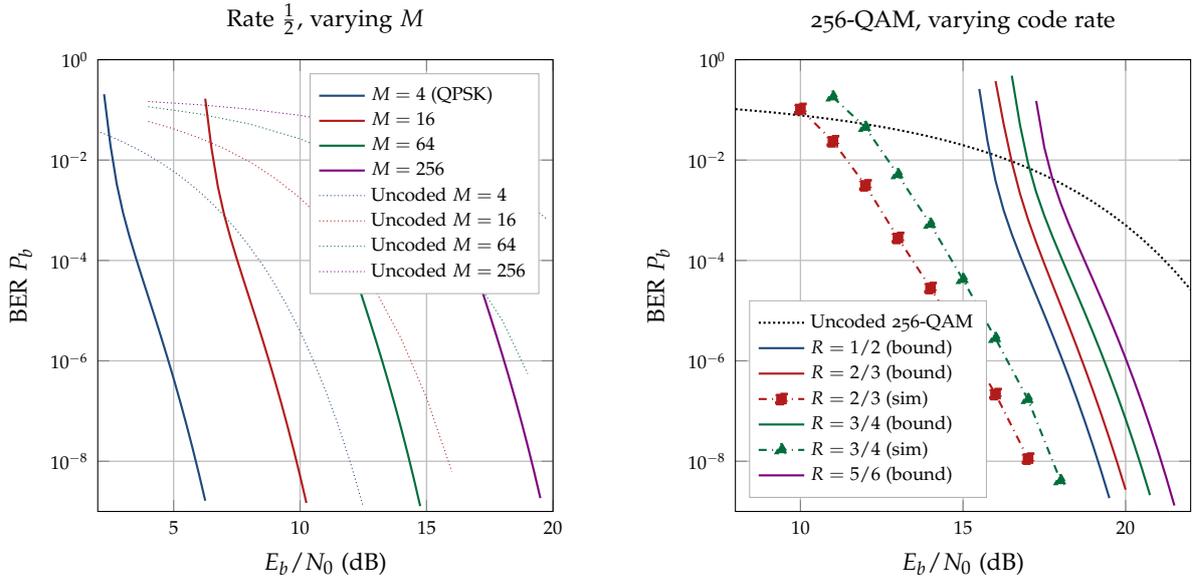
\begin{figure*}[ht]
\centering
\begin{tikzpicture}
\begin{groupplot}[
  group style={
    group size=2 by 1,
    horizontal sep=2.4cm,
  },
  ymode=log, ymin=1e-9, ymax=1,
  grid=both,
  grid style={line width=0.3pt, draw=gray!30},
  major grid style={line width=0.5pt, draw=gray!50},
  tick label style={font=\scriptsize},
  label style={font=\small},
  xlabel={$E_b/N_0$ (dB)},
  legend style={font=\scriptsize, at={(0.97,0.97)},
    anchor=north east, draw=gray!60, fill=white,
    inner sep=3pt, row sep=-1pt},
  legend cell align=left,
  clip=true,
]

%% ============================================================
%% LEFT PANEL — Rate 1/2, varying M
%% ============================================================
\nextgroupplot[
  width=0.46\textwidth, height=0.46\textwidth,
  xmin=2, xmax=20,
  ylabel={BER $P_b$},
  title={Rate $\frac{1}{2}$, varying $M$},
  title style={font=\small},
]

%% Coded BEP curves (solid)
\addplot[tufteblue, solid, thick] coordinates {
  (2.25,2.0530e-01) (2.50,1.9933e-02) (2.75,3.3325e-03) (3.00,8.8959e-04)
  (3.25,3.0325e-04) (3.50,1.1551e-04) (3.75,4.6191e-05) (4.00,1.8775e-05)
  (4.25,7.6079e-06) (4.50,3.0331e-06) (4.75,1.1784e-06) (5.00,4.4272e-07)
  (5.25,1.5984e-07) (5.50,5.5140e-08) (5.75,1.8082e-08) (6.00,5.6092e-09)
  (6.25,1.6379e-09)
};
\addlegendentry{$M=4$ (QPSK)}

\addplot[tuftered, solid, thick] coordinates {
  (6.25,1.6678e-01) (6.50,1.6826e-02) (6.75,2.9460e-03) (7.00,8.0864e-04)
  (7.25,2.7922e-04) (7.50,1.0697e-04) (7.75,4.2876e-05) (8.00,1.7434e-05)
  (8.25,7.0581e-06) (8.50,2.8088e-06) (8.75,1.0885e-06) (9.00,4.0772e-07)
  (9.25,1.4668e-07) (9.50,5.0400e-08) (9.75,1.6456e-08) (10.00,5.0800e-09)
  (10.25,1.4757e-09)
};
\addlegendentry{$M=16$}

\addplot[tuftegreen, solid, thick] coordinates {
  (10.75,1.2567e-01) (11.00,1.3407e-02) (11.25,2.4978e-03) (11.50,7.1064e-04)
  (11.75,2.4943e-04) (12.00,9.6261e-05) (12.25,3.8695e-05) (12.50,1.5739e-05)
  (12.75,6.3629e-06) (13.00,2.5256e-06) (13.25,9.7533e-07) (13.50,3.6376e-07)
  (13.75,1.3022e-07) (14.00,4.4497e-08) (14.25,1.4439e-08) (14.50,4.4280e-09)
  (14.75,1.2770e-09)
};
\addlegendentry{$M=64$}

\addplot[violet, solid, thick] coordinates {
  (15.50,2.6108e-01) (15.75,2.4319e-02) (16.00,3.8532e-03) (16.25,9.9423e-04)
  (16.50,3.3362e-04) (16.75,1.2617e-04) (17.00,5.0309e-05) (17.25,2.0438e-05)
  (17.50,8.2891e-06) (17.75,3.3113e-06) (18.00,1.2900e-06) (18.25,4.8634e-07)
  (18.50,1.7629e-07) (18.75,6.1092e-08) (19.00,2.0135e-08) (19.25,6.2799e-09)
  (19.50,1.8446e-09)
};
\addlegendentry{$M=256$}

%% Uncoded references (dotted, same colours)
\addplot[tufteblue, densely dotted, thin] coordinates {
  (0.00,7.8650e-02) (0.50,6.7065e-02) (1.00,5.6282e-02) (1.50,4.6401e-02)
  (2.00,3.7506e-02) (2.50,2.9655e-02) (3.00,2.2878e-02) (3.50,1.7173e-02)
  (4.00,1.2501e-02) (4.50,8.7938e-03) (5.00,5.9539e-03) (5.50,3.8622e-03)
  (6.00,2.3883e-03) (6.50,1.3998e-03) (7.00,7.7267e-04) (7.50,3.9880e-04)
  (8.00,1.9091e-04) (8.50,8.4000e-05) (9.00,3.3627e-05) (9.50,1.2109e-05)
  (10.00,3.8721e-06) (10.50,1.0838e-06) (11.00,2.6131e-07) (11.50,5.3287e-08)
  (12.00,9.0060e-09) (12.50,1.2330e-09)
};
\addlegendentry{Uncoded $M=4$}

\addplot[tuftered, densely dotted, thin] coordinates {
  (4.00,5.8618e-02) (5.00,4.1892e-02) (6.00,2.7871e-02) (7.00,1.6967e-02)
  (8.00,9.2472e-03) (9.00,4.3903e-03) (10.00,1.7542e-03) (11.00,5.6471e-04)
  (12.00,1.3866e-04) (13.00,2.4234e-05) (14.00,2.7632e-06) (15.00,1.8419e-07)
  (16.00,6.2502e-09)
};
\addlegendentry{Uncoded $M=16$}

\addplot[tuftegreen, densely dotted, thin] coordinates {
  (4.00,1.1576e-01) (6.00,8.3473e-02) (8.00,5.2320e-02) (10.00,2.6533e-02)
  (12.00,9.7240e-03) (14.00,2.1540e-03) (16.00,2.1717e-04) (18.00,6.3511e-06)
  (19.00,5.5374e-07)
};
\addlegendentry{Uncoded $M=64$}

\addplot[violet, densely dotted, thin] coordinates {
  (4.00,1.4691e-01) (6.00,1.2667e-01) (8.00,1.0334e-01) (10.00,7.7807e-02)
  (12.00,5.2022e-02) (14.00,2.9098e-02) (16.00,1.2400e-02) (18.00,3.4721e-03)
  (19.75,6.7378e-04)
};
\addlegendentry{Uncoded $M=256$}

%% ============================================================
%% RIGHT PANEL — 256-QAM, varying code rate  +  sim markers
%% ============================================================
\nextgroupplot[
  width=0.46\textwidth, height=0.46\textwidth,
  xmin=8, xmax=22,
  ylabel={BER $P_b$},
  title={256-QAM, varying code rate},
  title style={font=\small},
  legend style={font=\scriptsize, at={(0.03,0.03)},
    anchor=south west, draw=gray!60, fill=white,
    inner sep=3pt, row sep=-1pt},
]

%% Uncoded 256-QAM reference
\addplot[black, densely dotted, thick] coordinates {
  (8.00,1.0334e-01) (8.50,9.7112e-02) (9.00,9.0758e-02) (9.50,8.4312e-02)
  (10.00,7.7807e-02) (10.50,7.1279e-02) (11.00,6.4773e-02) (11.50,5.8337e-02)
  (12.00,5.2022e-02) (12.50,4.5883e-02) (13.00,3.9978e-02) (13.50,3.4365e-02)
  (14.00,2.9098e-02) (14.50,2.4230e-02) (15.00,1.9803e-02) (15.50,1.5853e-02)
  (16.00,1.2400e-02) (16.50,9.4516e-03) (17.00,6.9996e-03) (17.50,5.0194e-03)
  (18.00,3.4721e-03) (18.50,2.3070e-03) (19.00,1.4654e-03) (19.50,8.8513e-04)
  (20.00,5.0531e-04) (20.50,2.7082e-04) (21.00,1.3524e-04) (21.50,6.2388e-05)
  (22.00,2.6336e-05)
};
\addlegendentry{Uncoded 256-QAM}

%% ---- Rate 1/2: bound only (no sim data) ----
\addplot[tufteblue, solid, thick] coordinates {
  (15.50,2.6108e-01) (15.75,2.4319e-02) (16.00,3.8532e-03) (16.25,9.9423e-04)
  (16.50,3.3362e-04) (16.75,1.2617e-04) (17.00,5.0309e-05) (17.25,2.0438e-05)
  (17.50,8.2891e-06) (17.75,3.3113e-06) (18.00,1.2900e-06) (18.25,4.8634e-07)
  (18.50,1.7629e-07) (18.75,6.1092e-08) (19.00,2.0135e-08) (19.25,6.2799e-09)
  (19.50,1.8446e-09)
};
\addlegendentry{$R=1/2$ (bound)}

%% ---- Rate 2/3: bound + sim ----
\addplot[tuftered, solid, thick] coordinates {
  (16.00,3.7473e-01) (16.25,5.2178e-02) (16.50,9.1259e-03) (16.75,2.1612e-03)
  (17.00,6.5496e-04) (17.25,2.2909e-04) (17.50,8.5998e-05) (17.75,3.3262e-05)
  (18.00,1.2949e-05) (18.25,4.9998e-06) (18.50,1.8952e-06) (18.75,6.9992e-07)
  (19.00,2.5033e-07) (19.25,8.6271e-08) (19.50,2.8525e-08) (19.75,9.0131e-09)
  (20.00,2.7114e-09)
};
\addlegendentry{$R=2/3$ (bound)}
\addplot[tuftered, dashdotted, thick, mark=square*, mark size=2pt] coordinates {
  (10.00,1.0437e-01) (11.00,2.3658e-02) (12.00,3.1421e-03) (13.00,2.8133e-04)
  (14.00,2.8275e-05) (15.00,2.6202e-06) (16.00,2.1783e-07) (17.00,1.1284e-08)
};
\addlegendentry{$R=2/3$ (sim)}

%% ---- Rate 3/4: bound + sim ----
\addplot[tuftegreen, solid, thick] coordinates {
  (16.50,4.8050e-01) (16.75,5.1085e-02) (17.00,8.6520e-03) (17.25,2.2875e-03)
  (17.50,7.8097e-04) (17.75,3.0123e-04) (18.00,1.2297e-04) (18.25,5.1360e-05)
  (18.50,2.1509e-05) (18.75,8.9105e-06) (19.00,3.6152e-06) (19.25,1.4252e-06)
  (19.50,5.4241e-07) (19.75,1.9813e-07) (20.00,6.9097e-08) (20.25,2.2894e-08)
  (20.50,7.1721e-09) (20.75,2.1144e-09)
};
\addlegendentry{$R=3/4$ (bound)}
\addplot[tuftegreen, dashdotted, thick, mark=triangle*, mark size=2.5pt] coordinates {
  (11.00,1.7969e-01) (12.00,4.4492e-02) (13.00,5.0880e-03) (14.00,5.2224e-04)
  (15.00,4.1473e-05) (16.00,2.7771e-06) (17.00,1.6884e-07) (18.00,4.0697e-09)
};
\addlegendentry{$R=3/4$ (sim)}

%% ---- Rate 5/6: bound only (no sim data) ----
\addplot[violet, solid, thick] coordinates {
  (17.25,1.5092e-01) (17.50,1.7827e-02) (17.75,4.2027e-03) (18.00,1.3990e-03)
  (18.25,5.4065e-04) (18.50,2.2363e-04) (18.75,9.5146e-05) (19.00,4.0690e-05)
  (19.25,1.7227e-05) (19.50,7.1421e-06) (19.75,2.8750e-06) (20.00,1.1160e-06)
  (20.25,4.1520e-07) (20.50,1.4728e-07) (20.75,4.9563e-08) (21.00,1.5748e-08)
  (21.25,4.7027e-09) (21.50,1.3137e-09)
};
\addlegendentry{$R=5/6$ (bound)}

\end{groupplot}
\end{tikzpicture}
\caption{BEP union bounds~\eqref{eq:ub_qam} for Gray-coded $M$-QAM over AWGN
  with 30~spectrum terms ($d_{\max}=130$)~\cite{proakis01,lin04,caire98}.
  \textit{Left:} code rate~$\frac{1}{2}$ fixed; solid lines are coded bounds,
  dotted lines are uncoded $M$-QAM references.  Each factor-of-4 increase in
  $M$ shifts the waterfall $\approx$4--5\,dB to the right (see $\Delta_M$,
  Table~\ref{tab:deltaM}).
  \textit{Right:} 256-QAM fixed; all four code rate bounds are shown (solid),
  with Monte Carlo simulation markers ($\bullet$, dot-dashed) for $R=2/3$
  and $R=3/4$.  The uncoded 256-QAM reference is shown dotted.}
\label{fig:ber_qam}
\end{figure*}

%% =======================================================
\section{Augmented Trellis and Transfer Function}
%% =======================================================

\subsection{Why augment the state}

For the unpunctured code the transfer function method~\cite{lin04,viterbi79} uses a single
$64\times 64$ state space.  Puncturing makes the branch weight
\emph{phase-dependent}: consecutive encoder steps may contribute
differently depending on which bits survive the mask.  Augmenting the
state with the puncture phase converts this into a time-invariant problem~\cite{hagenauer88}:

\begin{definition}[Augmented state]
$(\sigma, \phi)$ where $\sigma\in\{0,\ldots,63\}$ is the encoder state
and $\phi\in\{0,\ldots,L-1\}$ is the position in the puncture mask.
\end{definition}

For rate-$\frac{3}{4}$ ($L=6$) the augmented space has $64\times6=384$ states.

\subsection{Branch weights}

Each branch carries weight $\D^{d}\N^{u}$ where
$d = p_\phi v^{(1)} + p_{(\phi+1)\bmod L} v^{(2)}$
counts transmitted 1-bits, and the phase advances by~2 per step.

\subsection{State partitioning}

Let $\mathcal{S}=\{(\sigma,\phi)\neq(0,0)\}$.  Branches split into:
\quad $\Emat$: \textsc{start}$\to\mathcal{S}$;\quad
$\Qmat$: $\mathcal{S}\to\mathcal{S}$;\quad
$\Rmat$: $\mathcal{S}\to$\textsc{start}.

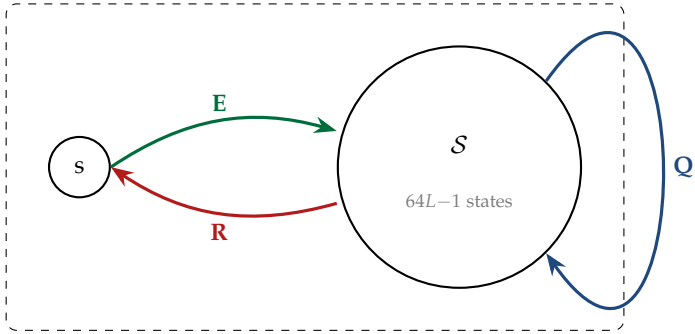
\begin{figure}[ht]
\centering
\begin{tikzpicture}[
  snode/.style  = {circle, draw, thick, minimum size=0.80cm, font=\small},
  bigcirc/.style= {circle, draw, thick, minimum size=3.2cm},
  >=Stealth, thick
]
  \node[snode] (start) at (0,0) {\textsc{s}};
  \node[bigcirc] (S) at (5,0) {};
  \node[font=\small]       at (5,  0.30) {$\mathcal{S}$};
  \node[font=\scriptsize, color=gray] at (5, -0.45) {$64L{-}1$ states};

  \draw[->, color=tuftegreen, bend left=25, very thick]
    (start.east) to node[above,font=\small]{$\Emat$} ($(S.west)+(0,0.48)$);
  \draw[->, color=tuftered, bend left=25, very thick]
    ($(S.west)+(0,-0.48)$) to node[below,font=\small]{$\Rmat$} (start.east);
  \draw[->, color=tufteblue, very thick]
    (S.north east) to[out=55,in=-55,looseness=4]
    node[right,font=\small]{$\Qmat$} (S.south east);

  \begin{scope}[on background layer]
    \node[draw, rounded corners, dashed, inner sep=0.55cm,
          fit=(start)(S), fill=white] {};
  \end{scope}
\end{tikzpicture}
\caption{$\Emat/\Qmat/\Rmat$ partition of the augmented trellis.
  The transfer function sums over all first-return paths from
  \textsc{start} to \textsc{start}.}
\label{fig:state-partition}
\end{figure}

%% =======================================================
\section{Computing the Transfer Function}
%% =======================================================

\subsection{First-return generating function}

The \emph{first-return transfer polynomial}~\cite{lin04,viterbi79,blahut83} is
\[
  T(\D,\N) = \Emat^\top \bigl(\mathbf{I}-\Qmat\bigr)^{-1} \Rmat,
  \qquad
  \alpha_d = [\D^d]\,T(\D,1),\quad
  \beta_d  = [\D^d]\,\tfrac{\partial T}{\partial \N}\big|_{\N=1}.
\]

\subsection{Neumann series (truncated)}

The Neumann (geometric) series identity~\cite{lin04,ryan09}
\[
  \bx = (\mathbf{I}-\Qmat)^{-1}\Rmat = \sum_{k\geq 0}\Qmat^k\Rmat
\]
avoids explicit matrix inversion.
Each $\Qmat$-multiplication raises the minimum degree by at least~1,
so the sum terminates within $d_{\max}$ steps.

\subsection{Implementation notes}

\begin{description}
  \item[Sparse polynomials.]
    \texttt{Dict[int $\to$ (count, weight)]}: only non-zero distances stored.
  \item[Combined multiply.]
    $(c_1,w_1)\cdot(c_2,w_2)=(c_1 c_2,\;w_1 c_2+w_2 c_1)$ tracks both
    $\alpha$ and $\beta$ in one pass.
  \item[Early termination.]
    Iteration halts when all polynomial terms exceed $d_{\max}$.
\end{description}

%% =======================================================
\section{Pseudocode}
%% =======================================================

\begin{figure*}[ht]
\begin{algorithm}[H]
\caption{First-Return Spectrum for Punctured BCC}
\begin{algorithmic}[1]
\Require Puncture mask $\mathbf{p}\in\{0,1\}^L$, max distance $d_{\max}$
\Ensure $d_\mathrm{free}$, $\{\alpha_d\}$, $\{\beta_d\}$ for $d\leq d_{\max}$
\For{each augmented state $(\sigma,\phi)$ and $u\in\{0,1\}$}
  \State Compute $(\sigma',\phi')$, outputs $(v^{(1)},v^{(2)})$ via \eqref{eq:state-update}--\eqref{eq:output}
  \State $d \leftarrow p_\phi v^{(1)} + p_{(\phi+1)\bmod L} v^{(2)}$;\quad
         $\mathrm{pol} \leftarrow \{d\mapsto(1,u)\}$
  \State Route pol to $\Emat$, $\Qmat$, or $\Rmat$ according to src/dst vs.\ \textsc{start}
\EndFor
\State $\bx\leftarrow\mathbf{0}$; $\mathrm{term}\leftarrow\Rmat$
\Repeat
  \State $\bx \mathrel{+}= \mathrm{term}$;\quad
         $\mathrm{term} \leftarrow \Qmat\cdot\mathrm{term}$ (sparse, truncate at $d_{\max}$)
\Until{$\mathrm{term}=\mathbf{0}$}
\State $T \leftarrow \Emat^\top\bx$;\quad remove $T[0]$
\State $d_\mathrm{free}\leftarrow\min(\mathrm{keys}(T))$;\quad
       $\alpha_d\leftarrow T[d].\mathrm{count}$;\quad
       $\beta_d\leftarrow T[d].\mathrm{weight}$
\end{algorithmic}
\end{algorithm}
\caption{All-zero self-loop (\textsc{start}$\to$\textsc{start}, $u=0$)
  is excluded so only non-trivial error events are counted.}
\label{fig:algorithm}
\end{figure*}

%% =======================================================
%% =======================================================
\section{Full Distance Spectrum Tables}
%% =======================================================

Tables~\ref{tab:spec12}--\ref{tab:spec56} list the distance
spectrum for each of the four IEEE~802.11 BCC rates, computed
at $d_{\max}=200$ by all three provided implementations.
The leading entries near $d_\mathrm{free}$ dominate the
union bound~\eqref{eq:ub} at practical SNRs~\cite{lin04,proakis01};
the rapidly growing tail is tabulated for completeness.

\begin{table*}[ht]
\centering\scriptsize\setlength{\tabcolsep}{6pt}
\renewcommand{\arraystretch}{1.06}
\begin{tabular}{@{} r r r @{}}
\toprule
\multicolumn{3}{c}{\textbf{Rate~$\frac{1}{2}$,\ \ $d_{\rm free}=10$}}\\
\cmidrule(lr){1-3}
$d$ & $\alpha_d$ & $\beta_d$ \\
\midrule
  10 & 11 & 36 \\
  12 & 38 & 211 \\
  14 & 193 & 1\,404 \\
  16 & 1\,331 & 11\,633 \\
  18 & 7\,275 & 77\,433 \\
  20 & 40\,406 & 502\,690 \\
  22 & 234\,969 & 3\,322\,763 \\
  24 & 1\,337\,714 & 21\,292\,910 \\
  26 & 7\,594\,819 & 134\,365\,911 \\
  28 & 43\,375\,588 & 843\,425\,871 \\
  30 & 247\,339\,453 & 5\,245\,283\,348 \\
  32 & 1\,409\,277\,901 & 32\,372\,937\,519 \\
  34 & 8\,034\,996\,288 & 198\,723\,833\,069 \\
  36 & 45\,808\,756\,116 & 1\,213\,657\,958\,889 \\
  38 & 261\,128\,775\,464 & 7\,378\,557\,447\,583 \\
  40 & 1\,488\,634\,502\,286 & 44\,686\,304\,667\,721 \\
  42 & 8\,486\,419\,243\,793 & 269\,700\,526\,164\,453 \\
  44 & 48\,378\,617\,913\,225 & 1\,622\,729\,997\,782\,985 \\
  46 & 275\,793\,790\,396\,626 & 9\,736\,706\,163\,099\,939 \\
  48 & 1\,572\,231\,420\,375\,534 & 58\,276\,780\,724\,903\,312 \\
  50 & 8\,962\,880\,896\,223\,488 & 348\,013\,358\,443\,369\,656 \\
  52 & 51\,095\,054\,431\,014\,717 & 2\,073\,963\,288\,833\,171\,453 \\
  54 & 291\,279\,733\,217\,405\,452 & 12\,336\,364\,663\,874\,535\,729 \\
  56 & 1\,660\,510\,362\,328\,999\,082 & 73\,252\,242\,728\,489\,893\,106 \\
  58 & 9\,466\,139\,591\,359\,800\,558 & 434\,271\,414\,857\,187\,001\,700 \\
  60 & 53\,964\,013\,243\,208\,867\,416 & 2\,570\,753\,624\,392\,467\,445\,784 \\
  62 & 307\,634\,876\,470\,723\,955\,762 & 15\,197\,254\,111\,652\,957\,956\,012 \\
  64 & 1\,753\,746\,820\,351\,580\,399\,624 & 89\,725\,716\,153\,372\,815\,432\,783 \\
  66 & 9\,997\,656\,840\,401\,781\,321\,694 & 529\,118\,969\,577\,464\,553\,987\,962 \\
  68 & 56\,994\,054\,713\,893\,565\,476\,484 & 3\,116\,793\,898\,985\,013\,529\,095\,287 \\
  70 & 324\,908\,358\,143\,739\,422\,211\,879 & 18\,340\,524\,278\,975\,143\,091\,254\,596 \\
  72 & 1\,852\,218\,477\,505\,497\,507\,791\,900 & 107\,818\,193\,462\,006\,470\,300\,574\,577 \\
  74 & 10\,559\,018\,266\,445\,912\,586\,742\,730 & 633\,248\,620\,007\,255\,756\,144\,580\,769 \\
  76 & 60\,194\,230\,912\,459\,137\,717\,773\,409 & 3\,716\,048\,802\,987\,112\,205\,321\,093\,900 \\
  78 & 343\,151\,734\,741\,456\,193\,472\,066\,807 & 21\,788\,866\,274\,369\,339\,795\,691\,938\,329 \\
  80 & 1\,956\,219\,246\,813\,370\,209\,261\,798\,213 & 127\,659\,548\,322\,985\,097\,263\,197\,345\,800 \\
  82 & 11\,151\,899\,740\,282\,547\,158\,768\,192\,994 & 747\,403\,707\,510\,664\,069\,605\,252\,957\,557 \\
  84 & 63\,574\,094\,785\,214\,431\,866\,818\,170\,525 & 4\,372\,772\,741\,086\,158\,690\,229\,034\,611\,083 \\
  86 & 362\,419\,464\,118\,062\,623\,197\,028\,805\,210 & 25\,566\,629\,266\,561\,432\,354\,938\,081\,678\,201 \\
  88 & 2\,066\,059\,586\,301\,199\,607\,386\,445\,446\,232 & 149\,389\,136\,373\,987\,711\,925\,267\,978\,784\,505 \\
\bottomrule
\end{tabular}
\caption{Distance spectrum for rate~$\frac{1}{2}$, first~40~non-zero terms.  Only even-$d$ terms appear owing to the odd--even distance structure of the unpunctured code~\cite{lin04,viterbi79}.  Both $\alpha_d$ (event multiplicity) and $\beta_d$ (total  input-bit weight) grow exponentially with~$d$.  The first few rows near $d_{\rm free}$ dominate the  union-bound BEP~\protect\eqref{eq:ub}.}
\label{tab:spec12}
\end{table*}

\begin{table*}[ht]
\centering\scriptsize\setlength{\tabcolsep}{6pt}
\renewcommand{\arraystretch}{1.06}
\begin{tabular}{@{} r r r @{}}
\toprule
\multicolumn{3}{c}{\textbf{Rate~$\frac{2}{3}$,\ \ $d_{\rm free}=6$}}\\
\cmidrule(lr){1-3}
$d$ & $\alpha_d$ & $\beta_d$ \\
\midrule
  6 & 1 & 3 \\
  7 & 16 & 70 \\
  8 & 48 & 285 \\
  9 & 158 & 1\,276 \\
  10 & 642 & 6\,160 \\
  11 & 2\,435 & 27\,128 \\
  12 & 9\,174 & 117\,019 \\
  13 & 34\,705 & 498\,860 \\
  14 & 131\,585 & 2\,103\,891 \\
  15 & 499\,608 & 8\,784\,123 \\
  16 & 1\,893\,179 & 36\,328\,084 \\
  17 & 7\,172\,729 & 149\,215\,136 \\
  18 & 27\,191\,646 & 609\,374\,214 \\
  19 & 103\,077\,011 & 2\,475\,565\,587 \\
  20 & 390\,696\,502 & 10\,011\,487\,814 \\
  21 & 1\,480\,891\,596 & 40\,328\,889\,729 \\
  22 & 5\,613\,272\,624 & 161\,890\,464\,724 \\
  23 & 21\,276\,960\,168 & 647\,849\,333\,879 \\
  24 & 80\,649\,275\,876 & 2\,585\,310\,552\,363 \\
  25 & 305\,696\,805\,990 & 10\,290\,999\,621\,644 \\
  26 & 1\,158\,729\,619\,748 & 40\,870\,598\,781\,266 \\
  27 & 4\,392\,111\,993\,691 & 161\,979\,866\,397\,621 \\
  28 & 16\,648\,093\,424\,227 & 640\,744\,298\,836\,396 \\
  29 & 63\,103\,811\,287\,025 & 2\,530\,171\,149\,917\,233 \\
  30 & 239\,192\,038\,606\,689 & 9\,975\,074\,654\,637\,651 \\
  31 & 906\,646\,220\,176\,741 & 39\,267\,783\,332\,904\,885 \\
  32 & 3\,436\,599\,953\,926\,219 & 154\,368\,148\,990\,251\,773 \\
  33 & 13\,026\,270\,800\,979\,961 & 606\,069\,217\,780\,505\,410 \\
  34 & 49\,375\,468\,291\,633\,374 & 2\,376\,664\,159\,970\,176\,069 \\
  35 & 187\,155\,396\,191\,213\,743 & 9\,309\,547\,543\,710\,430\,089 \\
  36 & 709\,403\,747\,993\,606\,556 & 36\,428\,001\,853\,950\,581\,850 \\
  37 & 2\,688\,961\,620\,804\,198\,202 & 142\,402\,017\,658\,290\,454\,051 \\
  38 & 10\,192\,382\,858\,882\,494\,599 & 556\,155\,757\,695\,350\,210\,952 \\
  39 & 38\,633\,749\,008\,587\,856\,293 & 2\,170\,198\,700\,519\,724\,466\,025 \\
  40 & 146\,439\,412\,940\,381\,687\,661 & 8\,461\,485\,632\,028\,557\,640\,931 \\
  41 & 555\,071\,723\,869\,022\,381\,011 & 32\,965\,323\,736\,976\,789\,411\,788 \\
  42 & 2\,103\,973\,325\,619\,741\,975\,380 & 128\,336\,338\,609\,792\,333\,679\,995 \\
  43 & 7\,975\,012\,173\,381\,708\,369\,090 & 499\,275\,312\,124\,292\,023\,982\,821 \\
  44 & 30\,228\,909\,459\,430\,820\,775\,734 & 1\,941\,082\,510\,012\,205\,927\,466\,204 \\
  45 & 114\,581\,262\,979\,693\,073\,017\,508 & 7\,541\,809\,059\,515\,936\,836\,327\,070 \\
\bottomrule
\end{tabular}
\caption{Distance spectrum for rate~$\frac{2}{3}$, first~40~non-zero terms.  Puncturing one bit per period reduces $d_\mathrm{free}$ from~10 to~6 and introduces odd-$d$ terms~\cite{cain79,yasuda84}.  Both $\alpha_d$ (event multiplicity) and $\beta_d$ (total  input-bit weight) grow exponentially with~$d$.  The first few rows near $d_{\rm free}$ dominate the  union-bound BEP~\protect\eqref{eq:ub}.}
\label{tab:spec23}
\end{table*}

\begin{table*}[ht]
\centering\scriptsize\setlength{\tabcolsep}{6pt}
\renewcommand{\arraystretch}{1.06}
\begin{tabular}{@{} r r r @{}}
\toprule
\multicolumn{3}{c}{\textbf{Rate~$\frac{3}{4}$,\ \ $d_{\rm free}=5$}}\\
\cmidrule(lr){1-3}
$d$ & $\alpha_d$ & $\beta_d$ \\
\midrule
  5 & 8 & 42 \\
  6 & 31 & 201 \\
  7 & 160 & 1\,492 \\
  8 & 892 & 10\,469 \\
  9 & 4\,512 & 62\,935 \\
  10 & 23\,307 & 379\,644 \\
  11 & 121\,077 & 2\,253\,373 \\
  12 & 625\,059 & 13\,073\,811 \\
  13 & 3\,234\,886 & 75\,152\,755 \\
  14 & 16\,753\,077 & 428\,005\,675 \\
  15 & 86\,686\,071 & 2\,415\,121\,123 \\
  16 & 448\,565\,858 & 13\,534\,984\,705 \\
  17 & 2\,321\,546\,552 & 75\,422\,690\,722 \\
  18 & 12\,014\,661\,684 & 418\,134\,779\,192 \\
  19 & 62\,177\,678\,298 & 2\,307\,775\,877\,171 \\
  20 & 321\,782\,203\,428 & 12\,687\,767\,739\,589 \\
  21 & 1\,665\,294\,549\,473 & 69\,515\,274\,896\,547 \\
  22 & 8\,618\,250\,200\,425 & 379\,697\,527\,047\,278 \\
  23 & 44\,601\,241\,330\,678 & 2\,068\,214\,528\,915\,872 \\
  24 & 230\,820\,838\,592\,718 & 11\,237\,531\,722\,373\,744 \\
  25 & 1\,194\,546\,586\,395\,172 & 60\,920\,601\,474\,530\,625 \\
  26 & 6\,182\,030\,185\,381\,023 & 329\,581\,239\,552\,578\,272 \\
  27 & 31\,993\,308\,832\,099\,435 & 1\,779\,680\,104\,530\,834\,762 \\
  28 & 165\,572\,117\,751\,393\,610 & 9\,593\,328\,859\,682\,938\,761 \\
  29 & 856\,870\,607\,034\,189\,819 & 51\,630\,169\,441\,762\,663\,061 \\
  30 & 4\,434\,485\,984\,875\,218\,930 & 277\,457\,746\,783\,477\,047\,589 \\
  31 & 22\,949\,399\,628\,811\,599\,853 & 1\,489\,003\,802\,780\,537\,995\,255 \\
  32 & 118\,767\,980\,116\,369\,449\,256 & 7\,980\,719\,884\,803\,595\,786\,399 \\
  33 & 614\,649\,329\,566\,088\,851\,154 & 42\,724\,114\,238\,398\,358\,327\,533 \\
  34 & 3\,180\,939\,829\,090\,587\,663\,771 & 228\,466\,515\,787\,057\,620\,239\,973 \\
  35 & 16\,462\,034\,057\,400\,611\,109\,436 & 1\,220\,453\,061\,059\,539\,812\,179\,628 \\
  36 & 85\,194\,495\,923\,024\,617\,552\,504 & 6\,513\,229\,164\,273\,472\,492\,588\,230 \\
  37 & 440\,899\,472\,705\,962\,824\,580\,787 & 34\,727\,509\,242\,517\,223\,649\,661\,501 \\
  38 & 2\,281\,747\,698\,919\,298\,191\,628\,292 & 185\,001\,803\,203\,789\,088\,989\,606\,300 \\
  39 & 11\,808\,525\,262\,158\,996\,793\,797\,400 & 984\,746\,488\,048\,876\,648\,491\,885\,530 \\
  40 & 61\,111\,607\,095\,328\,770\,658\,831\,016 & 5\,237\,673\,761\,317\,170\,635\,847\,174\,896 \\
  41 & 316\,265\,447\,112\,328\,542\,272\,132\,605 & 27\,837\,856\,663\,491\,539\,147\,883\,618\,742 \\
  42 & 1\,636\,740\,347\,560\,052\,011\,522\,600\,685 & 147\,853\,948\,302\,437\,837\,631\,540\,003\,889 \\
  43 & 8\,470\,476\,271\,723\,617\,072\,730\,452\,262 & 784\,774\,762\,490\,643\,725\,716\,187\,973\,867 \\
  44 & 43\,836\,500\,014\,672\,964\,072\,435\,134\,751 & 4\,162\,805\,915\,963\,836\,391\,011\,696\,811\,539 \\
\bottomrule
\end{tabular}
\caption{Distance spectrum for rate~$\frac{3}{4}$, first~40~non-zero terms.  Two output bits per period are deleted; $d_\mathrm{free}$ falls to~5~\cite{yasuda84}.  Both $\alpha_d$ (event multiplicity) and $\beta_d$ (total  input-bit weight) grow exponentially with~$d$.  The first few rows near $d_{\rm free}$ dominate the  union-bound BEP~\protect\eqref{eq:ub}.}
\label{tab:spec34}
\end{table*}

\begin{table*}[ht]
\centering\scriptsize\setlength{\tabcolsep}{6pt}
\renewcommand{\arraystretch}{1.06}
\begin{tabular}{@{} r r r @{}}
\toprule
\multicolumn{3}{c}{\textbf{Rate~$\frac{5}{6}$,\ \ $d_{\rm free}=4$}}\\
\cmidrule(lr){1-3}
$d$ & $\alpha_d$ & $\beta_d$ \\
\midrule
  4 & 14 & 92 \\
  5 & 69 & 528 \\
  6 & 654 & 8\,694 \\
  7 & 4\,996 & 79\,453 \\
  8 & 39\,699 & 792\,114 \\
  9 & 315\,371 & 7\,375\,573 \\
  10 & 2\,507\,890 & 67\,884\,974 \\
  11 & 19\,921\,920 & 610\,875\,423 \\
  12 & 158\,275\,483 & 5\,427\,275\,376 \\
  13 & 1\,257\,455\,600 & 47\,664\,215\,639 \\
  14 & 9\,990\,453\,938 & 414\,847\,451\,604 \\
  15 & 79\,372\,452\,075 & 3\,583\,040\,670\,062 \\
  16 & 630\,602\,872\,400 & 30\,748\,409\,619\,146 \\
  17 & 5\,010\,053\,531\,956 & 262\,418\,568\,123\,539 \\
  18 & 39\,804\,179\,617\,382 & 2\,228\,895\,012\,849\,046 \\
  19 & 316\,238\,637\,713\,112 & 18\,852\,439\,937\,923\,866 \\
  20 & 2\,512\,471\,862\,922\,901 & 158\,870\,376\,816\,716\,859 \\
  21 & 19\,961\,238\,706\,464\,034 & 1\,334\,424\,696\,449\,318\,311 \\
  22 & 158\,589\,257\,850\,835\,062 & 11\,175\,609\,534\,518\,649\,264 \\
  23 & 1\,259\,969\,536\,714\,370\,401 & 93\,347\,351\,761\,800\,157\,709 \\
  24 & 10\,010\,282\,258\,845\,090\,282 & 777\,849\,869\,010\,260\,843\,014 \\
  25 & 79\,530\,296\,563\,407\,917\,449 & 6\,467\,657\,099\,669\,465\,163\,457 \\
  26 & 631\,857\,115\,254\,729\,072\,678 & 53\,670\,754\,342\,729\,722\,819\,253 \\
  27 & 5\,020\,016\,664\,677\,983\,514\,074 & 444\,569\,568\,207\,478\,822\,347\,564 \\
  28 & 39\,883\,332\,331\,658\,123\,789\,262 & 3\,676\,345\,005\,910\,658\,028\,271\,387 \\
\bottomrule
\end{tabular}
\caption{Distance spectrum for rate~$\frac{5}{6}$, first~25~non-zero terms.  Four output bits per period are deleted; $d_\mathrm{free}=4$ is the lowest of the four standard rates~\cite{hagenauer88,ieee80211}.  Both $\alpha_d$ (event multiplicity) and $\beta_d$ (total  input-bit weight) grow exponentially with~$d$.  The first few rows near $d_{\rm free}$ dominate the  union-bound BEP~\protect\eqref{eq:ub}.}
\label{tab:spec56}
\end{table*}

%% =======================================================
\section{Complexity and Correctness}
%% =======================================================

\paragraph{State space.}
$64L$ augmented states~\cite{hagenauer88}; for $L\leq10$ the matrix is at
most $639\times639$.  The Neumann iteration converges in milliseconds on
any modern workstation.

\paragraph{Convergence.}
$\Qmat^k\Rmat$ has minimum polynomial degree $\geq k$ (the all-zero path
is excluded by dropping the \textsc{start} self-loop~\cite{lin04,viterbi79}),
guaranteeing termination within $d_{\max}$ steps.

\paragraph{Validation.}
Published reference values~\cite{ieee80211,proakis01}
($d_\mathrm{free}=10$, $\alpha_{10}=11$, $\beta_{10}=36$ for
rate-$\frac{1}{2}$) are reproduced exactly by all three implementations.

%% =======================================================
%% =======================================================
\appendix
\section{Method: Augmented-Trellis First-Return Series}
\label{app:method}
%% =======================================================

The full descriptive title of the technique used throughout this report is
the \textbf{augmented-trellis first-return series expansion}.
``Augmented trellis'': the encoder state is extended by a puncture-phase
coordinate~\cite{hagenauer88,yasuda84}, converting a phase-dependent problem
into a time-invariant graph.
``First-return'': only paths from \textsc{start} back to \textsc{start}
(without revisiting it) are enumerated---exactly the set of error events
entering the union bound of Viterbi~\cite{viterbi67}.
``Series expansion'': the Neumann series $(I-Q)^{-1}=\sum_k Q^k$
avoids explicit matrix inversion~\cite{lin04,blahut83}.

This is closely related to the \emph{transfer function} or \emph{state
diagram} method described in Lin \& Costello~\cite{lin04}, Viterbi \&
Omura~\cite{viterbi79}, and Blahut~\cite{blahut83}; and to the
\emph{generating function} approach in Ryan \& Lin~\cite{ryan09}.
The phase augmentation for punctured codes is due to Hagenauer~\cite{hagenauer88}
and Yasuda et~al.~\cite{yasuda84}; the free-distance optimality of the
$\oct{133}/\oct{171}$ generators was established by
Odenwalder~\cite{odenwalder70} and documented for Viterbi-decoded
satellite systems by Heller \& Jacobs~\cite{heller71}.

\section{Spectrum Toolkit}
\label{app:impl}

Three independent implementations are provided, all producing bit-identical
results.  Source code and build instructions are available at
\url{https://github.com/geekymode/bcc\_spectrum}~\cite{bcc_github}.

\begin{description}
  \item[\texttt{python/bcc\_spectrum.py}]
    Pure Python~3, standard library only.  No external dependencies.
  \item[\texttt{julia/bcc\_spectrum.jl}]
    Julia with typed dictionaries and in-place polynomial arithmetic.
  \item[\texttt{cpp/bcc\_spectrum.cpp} + \texttt{bcc\_spectrum.h}]
    C++17 reusable library with a command-line driver in \texttt{main.cpp}.
\end{description}

\subsection*{Python}

\begin{tcolorbox}[codebox=pyblue, title={Python --- command line \& library usage}]
\begin{lstlisting}
# Run from terminal: d_max=30, show 5 terms per rate
python3 python/bcc_spectrum.py 30 5

# Library API
from python.bcc_spectrum import compute_spectrum, PUNCTURE_MASKS
dfree, alpha, beta = compute_spectrum(PUNCTURE_MASKS["1/2"], d_max=60)
print(dfree, alpha[10], beta[10])   # 10  11  36

# Custom puncture mask
my_mask = [1, 1, 0, 1, 1, 0]
dfree, alpha, beta = compute_spectrum(my_mask, d_max=80)
\end{lstlisting}
\end{tcolorbox}

\subsection*{Julia}

\begin{tcolorbox}[codebox=julgreen, title={Julia --- command line \& interactive usage}]
\begin{lstlisting}
# Run from terminal: d_max=30, show 5 terms per rate
julia julia/bcc_spectrum.jl 30 5

# REPL / script
include("julia/bcc_spectrum.jl")
dfree, alpha, beta = compute_spectrum(PUNCTURE_MASKS["3/4"], 60)
println(dfree, "  ", alpha[5], "  ", beta[5])   # 5  8  42
\end{lstlisting}
\end{tcolorbox}

\subsection*{C++}

\begin{tcolorbox}[codebox=cppred, title={C++17 --- build, run \& library usage}]
\begin{lstlisting}
# Build
g++ -std=c++17 -O2 -o bcc_spectrum cpp/bcc_spectrum.cpp cpp/main.cpp
# or: make

# Run from terminal: d_max=30, show 5 terms per rate
./bcc_spectrum 30 5

// Library API in your own code
#include "bcc_spectrum.h"
auto [dfree, alpha, beta] = bcc::compute_spectrum(bcc::mask_half(), 60);
// dfree=10  alpha[10]=11  beta[10]=36

auto [df23, a23, b23] = bcc::compute_spectrum(bcc::mask_two3(), 60);
\end{lstlisting}
\end{tcolorbox}

\subsection*{Sample output (\texttt{d\_max=30}, 5~terms per rate)}

\begin{tcolorbox}[codebox=outgray, title={Terminal output --- identical across all three implementations}]
\begin{lstlisting}
IEEE 802.11 BCC Distance Spectrum
K=7, generators 133_8 / 171_8
d_max=30, showing first 5 non-zero terms
============================================================

Rate 1/2  (puncture period = 2)
d_free = 10
      d                 alpha_d                    beta_d
  -----  ----------------------  ------------------------
     10                      11                        36
     12                      38                       211
     14                     193                     1,404
     16                   1,331                    11,633
     18                   7,275                    77,433

Rate 2/3  (puncture period = 4)
d_free = 6
      d                 alpha_d                    beta_d
  -----  ----------------------  ------------------------
      6                       1                         3
      7                      16                        70
      8                      48                       285
      9                     158                     1,276
     10                     642                     6,160

Rate 3/4  (puncture period = 6)
d_free = 5
      d                 alpha_d                    beta_d
  -----  ----------------------  ------------------------
      5                       8                        42
      6                      31                       201
      7                     160                     1,492
      8                     892                    10,469
      9                   4,512                    62,935

Rate 5/6  (puncture period = 10)
d_free = 4
      d                 alpha_d                    beta_d
  -----  ----------------------  ------------------------
      4                      14                        92
      5                      69                       528
      6                     654                     8,694
      7                   4,996                    79,453
      8                  39,699                   792,114
\end{lstlisting}
\end{tcolorbox}

%% =======================================================
\bibliographystyle{abbrvnat}
{\small\bibliography{bcc_spectrum}}

\end{document}